\documentclass[aip,jcp, reprint]{revtex4-1}

\draft

\usepackage{amsmath}
\usepackage{graphicx}
\usepackage{dcolumn}
\usepackage{braket}
\usepackage{enumitem}
\usepackage{xcolor}
\newcommand{\padd}[1]{\hat{a}_{#1}^{\dagger}}

\begin{document}
	\title{Method For Making 2-Electron Response Reduced Density Matrices Approximately $N$-Representable}
	
	\author{Caitlin Lanssens}
	\email{Caitlin.Lanssens@UGent.be}
	\affiliation{Center for Molecular Modeling, Ghent University, Technologiepark 903, 9052 Zwijnaarde, Belgium}
	\author{Paul W. Ayers}
	\affiliation{McMaster University, Department of Chemistry \& Chemical Biology, 1280 Main Street West, Hamilton, Ontario, Canada}
	\author{Dimitri Van Neck}
	\author{Stijn De Baerdemacker}
	\author{Klaas Gunst}
	\affiliation{Center for Molecular Modeling, Ghent University, Technologiepark 903, 9052 Zwijnaarde, Belgium}
	\author{Patrick Bultinck}
	\affiliation{Ghent University, Department of Inorganic and Physical Chemistry, Krijgslaan 281 (S3), 9000 Ghent, Belgium}
	
	\date{\today}
	
	\begin{abstract}
		In methods like geminal-based approaches or coupled cluster that are solved using the projected Schr\"{o}dinger equation, direct computation of the 2-electron reduced density matrix (2-RDM) is impractical and one falls back to a 2-RDM based on response theory. However, the 2-RDMs from response theory are not $N$-representable. That is, the response 2-RDM does not correspond to an actual physical $N$-electron wave function. We present a new algorithm for making these non-$N$-representable 2-RDMs approximately $N$-representable, \textit{i.e.} it has the right symmetry and normalization and it fulfills the $P$-, $Q$- and $G$-conditions. Next to an algorithm which can be applied to any 2-RDM, we have also developed a 2-RDM optimization procedure specifically for seniority-zero 2-RDMs. We aim to find the 2-RDM with the right properties that is the closest (in the sense of the Frobenius norm) to the non-$N$-representable 2-RDM by minimizing the square norm of the difference between this initial response 2-RDM and the targeted 2-RDM under the constraint that the trace is normalized and the 2-RDM, $Q$- and $G$-matrices are positive semidefinite, \textit{i.e.} their eigenvalues are non-negative. Our method is suitable for fixing non-$N$-representable 2-RDMs which are close to being $N$-representable. Through the $N$-representability optimization algorithm we add a small correction to the initial 2-RDM such that it fulfills the most important $N$-representability conditions.
	\end{abstract}

	\maketitle
	
	\section{Introduction} \label{sec:intro}
	In quantum mechanics, any system can be described by a wave function, given as a solution of the Schr\"{o}dinger equation.  However, in practice this equation is only exactly solvable for a few specific systems such as the hydrogen atom. For more complicated systems, we have to rely on approximate methods. The standard approach is to first perform a Hartree Fock\cite{helgaker2014molecular,szabo1989modern} (HF) calculation. Here, the wave function is that Slater determinant which gives the lowest possible energy. Methods like configuration interaction\cite{shavitt1977ci} (CI) and coupled cluster\cite{shavitt2009coupledcluster} (CC) theory improve upon HF by including an increasing number of Slater determinants which are classified according to their excitation level. The highest level of theory, full CI, is reached when all possible Slater determinants are included in the wave function expansion. However, the factorial scaling of full CI makes it prohibitively expensive for all but the smallest systems.
	
	The HF, CI and CC methods are often referred to as single reference because it is assumed that the wave function is dominated by a single electron configuration. However, whenever several electron configurations become (nearly-)degenerate, \textit{e.g.} during bond breaking processes, these methods fail. Such systems are said to be strongly correlated and a qualitatively accurate description of the system requires multiple Slater determinants. Different multi-reference approaches were developed to include these strong correlation effects, like \textit{e.g.} CASSCF,\cite{roos1980casscf} MRCI\cite{shavitt1998ci} and DMRG.\cite{white1992dmrg, white1999dmrg, garnetchan2011dmrg,wouters2014dmrg} However, due to their computational cost their applicability is also limited to small systems.
	
	Instead of classifying determinants according to their excitation level, determinants can also be described in terms of their seniority.\cite{ring2004seniority} The seniority of a determinant is the number of unpaired electrons. In this study, we use roman letters $a$, $b$, $\dots$ to denote spatial orbitals. We will only consider singlet pairing, so each spatial orbital can at most be occupied by two electrons with opposite spin. We denote the spin-up orbital and spin-down orbital by $a$ and $\bar{a}$, respectively. General spin orbitals are referred to by Greek letters $\alpha$, $\beta$, $\dots$
	
	The doubly occupied CI\cite{weinhold1967rdmI,weinhold1967rdmII} (DOCI) wave function is the most general seniority-zero wave function and only includes those Slater determinants where all spatial orbitals are doubly occupied. The DOCI method has gained a lot of interest in recent years because it was found to be an excellent model for describing static correlation\cite{bytautas2011seniority,kutzelnigg2012separation,limacher2013geminals,alcoba2014seniority,limacher2016apighierarchy} in strongly correlated systems. Although the number of determinants is strongly reduced compared to full CI, DOCI still scales combinatorially with the system size. Because the orbitals in DOCI space are either empty or doubly occupied, we introduce the pair creation and pair annihilation operators, $S_a^{\dagger} = \padd{a}\padd{\bar{a}}$ and $S_a = \hat{a}_{\bar{a}}\hat{a}_{a}$, where $\padd{a}$ and $\hat{a}_{a}$ are the single particle creation and annihilation operators from the second quantization formalism.\cite{helgaker2014molecular}
	
	Recently, several of us introduced a computationally tractable approximation to DOCI that is called the antisymmetric product of 1-reference orbital geminals\cite{limacher2013geminals,johnson2013geminals,johnson2015geminals,boguslawski2014geminals,boguslawski2014ap1rog,boguslawski2014nonvariational} (AP1roG), also known as pair-coupled cluster doubles\cite{stein2014ccd,henderson2014seniority,henderson2015pair,bulik2015can} (pCCD)
	\begin{equation}
	\ket{\Psi_{AP1roG}} = \prod_{i=1}^{N/2} \left(S_{i}^{\dagger} + \sum_{a=N/2+1}^{L} c_i^a S_a^{\dagger}\right) \ket{\theta}
	\end{equation} 
	with $N$ the even number of electrons, $L$ the number of spatial orbitals and $\ket{\theta}$ denotes the vacuum state with respect to geminal creation. The geminal coefficients ${c_i^a}$ link the geminals to the underlying single particle orbitals. To keep the computational cost low, the AP1roG energy is not determined variationally, but by using the projected Schr\"odinger equation (cfr. CC theory). A popular choice for AP1roG is to project against a fully paired reference determinant, $\bra{\Phi_0}$, and the set of its pair-excited determinants, $\bra{\Phi_i^a} = \bra{\Phi_0} S_i^{\dagger}S_a$, resulting in the following set of equations
	\begin{subequations}
		\begin{eqnarray}
			\braket{\Phi_0|\hat{H} - E|\Psi} &=& 0 \\
			\braket{\Phi_i^a|\hat{H} - E|\Psi} &=& 0
		\end{eqnarray}
	\end{subequations} 
	It has been shown that AP1roG is able to produce results which are nearly indistinguishable from DOCI at reasonable computational cost for molecular systems.\cite{limacher2013geminals,tecmer2014accuracy,tecmer2015actinide} It is important to note that DOCI, and also its approximations, are orbital dependent. Hence, orbital optimization is essential to obtain accurate results.\cite{limacher2014orbrot,boguslawski2014geminals,boguslawski2014ap1rog,boguslawski2014nonvariational,stein2014ccd,henderson2014seniority,henderson2015pair,shepherd2016pccd} 
	
	It is well-known that since electrons are indistinguishable and they only interact pairwise, the energy is in fact completely determined by the 2-electron reduced density matrix\cite{lowdin1955quantum,husimi1940some,mayer1955electron} (2-RDM). In contrast to the wave function which is dependent on $4N$ spin coordinates, the 2-RDM provides us with a much less complicated object to investigate a system. 
	
	A Hamiltonian with at most two-body interactions, can be written as
	\begin{equation}
	\hat{H} = \sum_{\alpha\beta}^{2L} h_{\alpha\beta}\padd{\alpha}\hat{a}_{\beta} + \frac{1}{2} \sum_{\alpha\beta\gamma\delta}^{2L} V_{\alpha\beta\gamma\delta}\padd{\alpha}\padd{\beta}\hat{a}_{\delta}\hat{a}_{\gamma}
	\end{equation}
	with $h_{\alpha\beta}$ and $V_{\alpha\beta\gamma\delta}$ the one- and two-electron matrix elements. The 1-RDM and 2-RDM are defined as the expectation values of the 1- and 2-electron reduced density operators,
	\begin{subequations}
	\begin{equation}
		\rho_{\alpha\beta} = \braket{\Psi|\padd{\alpha}\hat{a}_{\beta}|\Psi } 
		\label{1rdm}
	\end{equation}
		\begin{equation}
			\Gamma_{\alpha\beta\gamma\delta} = \braket{\Psi|\padd{\alpha}\padd{\beta}\hat{a}_{\delta}\hat{a}_{\gamma}|\Psi } 
			\label{2rdm}
		\end{equation}
	\end{subequations}
	Now, the ground state energy can be expressed as a function of the 2-RDM:
	\begin{equation}
	E_0 = \frac{1}{2}\sum_{\alpha\beta\gamma\delta}^{2L}\Gamma_{\alpha\beta\gamma\delta}\mathrm{K}_{\alpha\beta\gamma\delta}
	\end{equation}
	with $K$ the second order reduced Hamiltonian:
	\begin{equation}
	\mathrm{K}_{\alpha\beta\gamma\delta} = \frac{1}{N-1}\left(h_{\alpha\gamma}\delta_{\beta\delta} + h_{\beta\delta}\delta_{\alpha\gamma}\right) + V_{\alpha\beta\gamma\delta}
	\end{equation}
	
	The 2-RDM could be computed directly by evaluation of the expectation values, Eq.~(\ref{2rdm}). However, for non-variational methods like CC or AP1roG this can become computationally very expensive and it may be better to compute the 2-RDM matrix elements as the response of the energy to changes in the two-electron matrix elements
	\begin{equation} 
		\label{resp2dm}
		\Gamma_{\alpha\beta\gamma\delta} = \frac{\partial E}{\partial V_{\alpha\beta\gamma\delta}}
	\end{equation}
	According to the Hellman-Feynman theorem,\cite{helgaker2014molecular} for exact solutions of the Schr\"{o}dinger equation the response  2-RDM is equal to the definition Eq.~(\ref{2rdm}). However, when using non-variational methods where it is more practical to calculate the response 2-RDM, we will have to deal with the fact that in general this 2-RDM is not exact in the sense that it does not equal the definition in Eq.~(\ref{2rdm}). Moreover, in most cases these approximate 2-RDMs will not be $N$-representable.
	
	The $N$-representability problem was first described in detail by A.J. Coleman in 1963.\cite{coleman1963structure, coleman2000coulson, mazziotti2007advances} How do we know that a given 2-RDM is actually derivable from a physical (ensemble of) $N$-electron wave function(s)? The necessary and sufficient conditions for a 2-RDM to be $N$-representable are known,\cite{garrod1964reduction,kummer1967nrepresentability,coleman1977convex} however they are of no practical use since it would require the knowledge of the ground state energy of every possible two-particle Hamiltonian. In section \ref{sec:2rdm} we will recapitulate the necessary and sufficient conditions for $N$-representability on the 1-RDM, first derived by Coleman.\cite{coleman1963structure} Afterwards, we move on to the more complex problem of $N$-representability of the 2-RDM. The most important necessary conditions were originally proposed by Garrod and Percus.\cite{garrod1964reduction} Later, these so-called positivity conditions were generalized to a hierarchy of $N$-representability conditions.\cite{erdahl2000lower,mazziotti2001uncertainty} In the last part of this section we look at the 2-RDM derived from a seniority-zero wave function.\cite{weinhold1967rdmI,weinhold1967rdmII,poelmans2015v2dmdoci} In DOCI space, the structure of the 2-RDM is greatly simplified. Furthermore, the $N$-representability conditions can be reformulated. These DOCI $N$-representability conditions contain and extend on the diagonal conditions proposed by many authors.\cite{ayers2017advances,nakata2012diagonal}
	
	In section \ref{sec:algorithm} we propose an algorithm to make non-$N$-representable 2-RDMs approximately $N$-representable. While we focus on 2-RDMs resulting from response theory, the algorithm can in fact be used on any non-$N$-representable 2-RDM. The goal is to find the 2-RDM which is the closest to the initial 2-RDM but has the right symmetry, the correct trace, and satisfies the so-called $P$-, $Q$- and $G$-conditions. Next to the general $N$-representability optimization algorithm which can act on any 2-RDM, we have also implemented a procedure for the special case of 2-RDMs in seniority-zero space. Finally, some calculations illustrating the performance of both algorithms are given in section \ref{sec:results}.
	
	\section{$N$-representability}  \label{sec:2rdm}
	\subsection{General $N$-representability conditions}
	Some properties of the 1-RDM and 2-RDM can be directly derived from their definitions Eqs.~(\ref{1rdm}) and (\ref{2rdm}). They are both Hermitian and have the following symmetry properties:
	
	\begin{subequations}
		\begin{equation} 
			\label{sym1rdm}
			\rho_{\alpha\beta} = \rho_{\beta\alpha}
		\end{equation}
		\begin{equation} 
			\label{sym2rdm}
			\Gamma_{\alpha\beta\gamma\delta} = \Gamma_{\gamma\delta\alpha\beta}
		\end{equation}
		\begin{equation} 
			\label{antisym2rdm}
			\Gamma_{\alpha\beta\gamma\delta} = -\Gamma_{\beta\alpha\gamma\delta} = -\Gamma_{\alpha\beta\delta\gamma} = \Gamma_{\beta\alpha\delta\gamma}
		\end{equation}
	\end{subequations}
	Note that we assume all wave functions to be real and all density matrices to be real - symmetric. Both RDMs are normalized:
	\begin{subequations}
		\begin{equation} 
			\label{trace1rdm}
			\mathrm{Tr}(\rho_{}) = \sum_{\alpha}^{2L}\rho_{\alpha\alpha} = N
		\end{equation}
		\begin{equation} 
			\label{trace2rdm}
			\mathrm{Tr}(\Gamma) = \sum_{\alpha\beta}^{2L}\Gamma_{\alpha\beta\alpha\beta} = N(N-1)
		\end{equation}
	\end{subequations}
	Furthermore, the 1-RDM can be computed from the 2-RDM through contraction
	\begin{equation}
		\rho_{\alpha\gamma} = \frac{1}{N-1}\sum_{\beta}^{2L} \Gamma_{\alpha\beta\gamma\beta}
	\end{equation}
	These conditions are necessary, but they are far from sufficient to guarantee that a given 1-RDM or 2-RDM is actually $N$-representable. Additional conditions must be imposed in order to ensure $N$-representability.
	
	It was found that $p$-positivity conditions\cite{erdahlJin2000rdm, mazziotti2001uncertainty} for a $p$-RDM could be derived from the positive semidefinite property of a class of non-negative Hamiltonians of the form
	\begin{equation}
		\hat{H} = \hat{A}^{\dagger}\hat{A}
	\end{equation}   
	where $\hat{A}$ is a $p$-particle operator. The expectation value of this Hamiltonian must be non-negative, 
	\begin{equation} 
		\label{psd}
		\braket{\Psi|\hat{A}^{\dagger}\hat{A}|\Psi} \geq 0,
	\end{equation}
	\textit{i.e.} its matrix representation must be positive semidefinite, with different forms of this operator leading to different $p$-positivity conditions. A matrix $M$ is positive semidefinite if and only if all of its eigenvalues are greater than or equal to zero. This is denoted as $M \succeq 0$. 
	
	\subsubsection{$N$-representability of the 1-RDM\\}
	From the non-negativity of Eq.~(\ref{psd}) we can derive two 1-positivity conditions. By choosing either  $\hat{A}^{\dagger} = \sum_{\alpha}x_{\alpha}\padd{\alpha}$ or $\hat{A}^{\dagger} = \sum_{\alpha} x_{\alpha}\hat{a}_{\alpha}$ we get the $p$- and $q$-matrix with elements:
	\begin{subequations}
		\begin{eqnarray}
			p_{\alpha\beta} &=& \braket{\Psi|\padd{\alpha}\hat{a}_{\beta}|\Psi} = \rho_{\alpha\beta} \\ 
			q_{\alpha\beta} &=& \braket{\Psi|\hat{a}_{\alpha}\padd{\beta}|\Psi} \label{q1rdm}
		\end{eqnarray}	
	\end{subequations}
	These matrices must be positive semidefinite, \textit{i.e.} $p \succeq 0$ and $q \succeq 0$. The first condition is equivalent to stating that the 1-RDM is positive semidefinite and the second condition can be rewritten as a function of the 1-RDM by using the anticommutation relation $\{\hat{a}_{i}, \padd{j}\} = \delta_{ij}$. This leads to the following 1-positivity conditions:
	\begin{subequations}
		\begin{eqnarray}
			\rho_{} &\succeq& 0 \\
			I - \rho_{} &\succeq& 0
		\end{eqnarray}
	\end{subequations}
	These conditions force the eigenvalues of the 1-RDM to have values in the interval $[0,1]$. Coleman\cite{coleman1963structure} has proven that this is a necessary and sufficient condition and this means that 
	$N$-representability of the 1-RDM is completely determined by its eigenvalues. 
	
	\subsubsection{$N$-representability of the 2-RDM\\}
	While $N$-representability for the 1-RDM is quite easily enforced through its eigenvalues, the $N$-representability problem for the 2-RDM is much more complicated.\cite{liu2007QMA} Since the 2-RDM should be positive semidefinite, the lower bound on its eigenvalues is zero, moreover there also exists an upper bound on the eigenvalues of the 2-RDM.\cite{coleman1963structure}
	\begin{equation}
		0 \leq \lambda_{\Gamma} \leq
		\begin{cases}
			N - 1 \quad \,\text{  if $N$ is odd}\\
			N \quad\quad\quad \text{ if $N$ is even}
		\end{cases}
	\end{equation}
	with $\lambda_{\Gamma}$ obeying the eigenvalue equation
	\begin{equation}
		\sum_{\gamma\delta} \Gamma_{\alpha\beta\gamma\delta} X_{\gamma\delta} = \lambda_{\Gamma} X_{\alpha\beta}
	\end{equation}
	
	However, this alone is not sufficient to guarantee $N$-representability. Additional necessary conditions are derived by restricting the operator $\hat{A}^{\dagger}$ in Eq.~\eqref{psd} to the two-particle space. The resulting 2-positivity conditions\cite{garrod1964reduction,rosina1975variational,erdahl1978representability} impose positive semidefiniteness on the $P$-, $Q$- and $G$-matrices.
	\begin{itemize}
		\item $P$-condition: choosing $A^{\dagger} = \sum_{\alpha\beta}x_{\alpha\beta}\padd{\alpha}\padd{\beta}$ imposes the positive semidefiniteness of the $P$-matrix which is actually the 2-RDM
		\begin{equation} 
			\label{Pcond}
			P \succeq 0 \quad \mathrm{with} \quad P_{\alpha\beta\gamma\delta} = \braket{\Psi|\padd{\alpha}\padd{\beta}\hat{a}_{\delta}\hat{a}_{\gamma}|\Psi}
		\end{equation} 
		\item $Q$-condition: $A^{\dagger} = \sum_{\alpha\beta}x_{\alpha\beta}\hat{a}_{\alpha}\hat{a}_{\beta}$ leads to 
		\begin{equation} 
			\label{Qcond}
			Q \succeq 0 \quad \mathrm{with} \quad Q_{\alpha\beta\gamma\delta} = \braket{\Psi|\hat{a}_{\alpha}\hat{a}_{\beta}\padd{\delta}\padd{\gamma}|\Psi}
		\end{equation} 
		By rearranging the creation and annihilation operators using the same anticommutation relation as before the $Q$-matrix can be rewritten in terms of the 1-RDM and 2-RDM.
		\begin{eqnarray}
			Q_{\alpha\beta\gamma\delta}&=&\Gamma_{\alpha\beta\gamma\delta} + \delta_{\beta\delta}\delta_{\alpha\gamma} - \delta_{\alpha\delta}\delta_{\beta\gamma} - \delta_{\beta\delta}\rho_{\alpha\gamma} 
			\nonumber \\ &&  + \delta_{\alpha\delta}\rho_{\beta\gamma} + \delta_{\beta\gamma}\rho_{\alpha\delta} - \delta_{\alpha\gamma}\rho_{\beta\delta} \label{QfromP}
		\end{eqnarray}
		\item $G$-condition: choosing $A^{\dagger} = \sum_{\alpha\beta}x_{\alpha\beta}\padd{\alpha}\hat{a}_{\beta}$ results in
		\begin{equation} \label{Gcond}
		G \succeq 0 \quad \mathrm{with} \quad G_{\alpha\beta\gamma\delta} = \braket{\Psi|\padd{\alpha}\hat{a}_{\beta}\padd{\delta}\hat{a}_{\gamma}|\Psi}
		\end{equation}
		The $G$-matrix is also a function of the 1-RDM and 2-RDM
		\begin{equation} \label{GfromP}
		G_{\alpha\beta\gamma\delta} = \delta_{\beta\delta}\rho_{\alpha\gamma} - \Gamma_{\alpha\delta\gamma\beta}
		\end{equation}
	\end{itemize}
	It should be clear that while the $Q$-matrix has the same symmetry properties as the 2-RDM,
	the $G$-matrix is only symmetric upon exchange of the index pairs but loses the antisymmetry property upon interchanging indices within a pair.
	The $Q$-matrix is normalized as:
	\begin{equation} 
		\label{traceq}
		\mathrm{Tr}(Q)= (2L-N)(2L-N-1)
	\end{equation}
	$2L-N$ is the number of virtual orbitals or holes. This is in complete agreement with the interpretation of the 2-RDM as a 2-particle reduced density matrix and the $Q$-matrix as a 2-hole reduced density matrix. The $G$-matrix, a particle-hole RDM, is normalized by:
	\begin{equation} 
		\label{traceg}
		\mathrm{Tr}(G)= N(2L-N +1)
	\end{equation} 
	
	\subsection{$N$-representability conditions in seniority-zero space}
	In seniority-zero space, the spatial orbitals are either empty or doubly occupied. Both the structure of the 1-RDM and 2-RDM are greatly simplified for DOCI wave functions because the operator cannot change the number of broken pairs, \textit{i.e.} the operators need to couple to seniority-zero.
	\subsubsection{$N$-representability of the 1-RDM\\}
	The 1-RDM becomes diagonal
	\begin{equation}
	\gamma_{ab} = \braket{\Psi|\padd{a}\hat{a}_{b}|\Psi} = \braket{\Psi|\padd{\bar{a}}\hat{a}_{\bar{b}}|\Psi} = \delta_{ab}\rho_{a}
	\end{equation}
	We have the following conditions:
	\begin{subequations}
		\begin{equation}
			\rho_{a} \succeq 0
		\end{equation}
		\begin{equation}
			\textrm{Tr}(\rho_{}) = \frac{N}{2}
		\end{equation}
	\end{subequations}
	The necessary and sufficient condition for $N$-representability of the 1-RDM again forces the elements of  $\rho_{a}$ to be in the interval $[0, 1]$.
	\subsubsection{$N$-representability of the 2-RDM\\} \label{sec:dociNrepconditions}
	Since seniority must be conserved, the 2-RDM becomes block diagonal in the seniority number and the $P$-, $Q$- and $G$-conditions can be reformulated.\cite{weinhold1967rdmI,weinhold1967rdmII,poelmans2015v2dmdoci}
	\begin{itemize}
		\item $P$-condition: the seniority-zero block of the 2-RDM is called the pair density matrix $\Pi$ with elements
		\begin{equation} \label{doci_pair}
			\forall a,b: \quad \Pi_{ab} = \braket{\Psi|\padd{a}\padd{\bar{a}}\hat{a}_{\bar{b}}\hat{a}_{b}|\Psi} = \Gamma_{a\bar{a}b\bar{b}}
		\end{equation}
		From the non-negativity of the Hamiltonian $\hat{A}^{\dagger}\hat{A}$ (Eq.~(\ref{psd})) with $\hat{A}^{\dagger} = \sum_a x_a \padd{a}\padd{\bar{a}}$ it follows that this pair density matrix should be positive semidefinite, \textit{i.e.} $\Pi \succeq 0$. The seniority-two block is part of the diagonal of the 2-RDM and we call it the exchange 2-RDM.
	\begin{subequations} \label{doci_exchange}
		\begin{eqnarray}
			\forall a\neq b: \quad D_{ab} &=& \braket{\Psi|\padd{a}\padd{b}\hat{a}_{b}\hat{a}_{a}|\Psi} = \Gamma_{abab} \\
			&=& \braket{\Psi|\padd{a}\padd{\bar{b}}\hat{a}_{\bar{b}}\hat{a}_{a}|\Psi} = \Gamma_{a\bar{b}a\bar{b}} \\
			&=& \braket{\Psi|\padd{\bar{a}}\padd{b}\hat{a}_{b}\hat{a}_{\bar{a}}|\Psi} = \Gamma_{\bar{a}b\bar{a}b} \\
			&=& \braket{\Psi|\padd{\bar{a}}\padd{\bar{b}}\hat{a}_{\bar{a}}\hat{a}_{\bar{b}}|\Psi} = \Gamma_{\bar{a}\bar{b}\bar{a}\bar{b}}
		\end{eqnarray}
	\end{subequations}
		We choose to put $D_{aa} = 0$. Because the diagonal elements of a positive semidefinite matrix must be non-negative, this leads to the condition that $D_{ab} \geq 0$. Now there are two ways of obtaining the 1-RDM from the 2-RDM
		\begin{subequations}
			\begin{eqnarray}
				\rho_{a} &=& \Pi_{aa} \\
				\rho_{a} &=& \frac{1}{\frac{N}{2} - 1}\sum_b^L D_{ab}
			\end{eqnarray}
		\end{subequations}
		The consistency between these two must be separately enforced. The trace of the 2-RDM can now be split into two contributions
		\begin{subequations}
			\begin{eqnarray}
				\mathrm{Tr}(\Pi) &=& \sum_a^L \Pi_{aa} = \frac{N}{2} \label{tracepair2rdm}\\
				\mathrm{Tr}(D) &=& \sum_{ab}^L D_{ab} = \frac{N}{2}\left(\frac{N}{2}-1\right) \label{traceexch2rdm}	
			\end{eqnarray}
		\end{subequations}
		
		\item $Q$-condition: the $Q$-matrix had the same structure as the 2-RDM. The seniority-zero part has the elements
		\begin{subequations}
			\begin{eqnarray}
				\forall a,b: \quad Q^{\Pi}_{ab} &=& \braket{\Psi|\hat{a}_{a}\hat{a}_{\bar{a}}\padd{\bar{b}}\padd{b}|\Psi} \\
				&=& \delta_{ab} \left(1- 2\Pi_{aa} \right) + \Pi_{ab}
			\end{eqnarray}
		\end{subequations}
		The constraint $Q^{\Pi} \succeq 0 $ follows from the non-negativity of Eq.~(\ref{psd}) when choosing $\hat{A}^{\dagger} = \sum_{a} x_a \hat{a}_{a}\hat{a}_{\bar{a}}$. 
		The seniority-two block is on the diagonal of the $Q$-matrix and all elements should be equal to or greater than zero
		\begin{subequations}
			\begin{eqnarray}
				\forall a \neq b: \quad Q^D_{ab} &=&\braket{\Psi|\hat{a}_{a}\hat{a}_{b}\padd{b}\padd{a}|\Psi}\\ 
				&=& D_{ab} + 1 - \Pi_{aa} - \Pi_{bb} \geq 0
		\end{eqnarray}
		\end{subequations}
		Their normalization is given by
		\begin{subequations}
			\begin{eqnarray}
				\mathrm{Tr}(Q^{\Pi}) &=& L - \frac{N}{2} \\
				\mathrm{Tr} (Q^D) &=& \left(L - \frac{N}{2}\right)\left(L - \frac{N}{2} -1\right)
			\end{eqnarray}
		\end{subequations}
		
		\item $G$-condition: this condition is more elaborate because more contributions are non-zero. For a complete derivation of the $G$-condition in DOCI space, we refer to Ref.~\onlinecite{poelmans2015v2dmdoci}. Firstly, the matrix $G^{\Pi}$ with elements
		\begin{subequations}
			\begin{eqnarray}
				G^{\Pi}_{ab} &=& \braket{\Psi|\padd{a}\hat{a}_{a}\padd{b}\hat{a}_{b}|\Psi} \\
				&=& \braket{\Psi|\padd{a}\hat{a}_{a}\padd{\bar{b}}\hat{a}_{\bar{b}}|\Psi} \\
				&=& D_{ab} + \delta_{ab}\Pi_{aa}
			\end{eqnarray}
		\end{subequations}
		must be positive semidefinite, $G^{\Pi} \succeq 0$, and secondly, the following set of $2 \times 2$ matrices must be positive semidefinite.
		\begin{equation}
			\forall a < b: \quad G^{(2x2)} = \begin{bmatrix}
			\Pi_{aa} - D_{ab} & \Pi_{ab} \\ \Pi_{ab} & \Pi_{bb} - D_{ab}
			\end{bmatrix} \succeq 0
		\end{equation}
	\end{itemize}
	The DOCI energy can be split in two contributions from the pair density matrix $\Pi$ and the exchange density matrix $D$:
	\begin{equation}
	E_{DOCI} = \sum_{ab}^L\left(K^{\Pi}_{ab}\Pi_{ab} + K^D_{ab}D_{ab}\right) \label{doci_energy}
	\end{equation}
	where the DOCI reduced Hamiltonian gets a similar structure as the 2-RDM does in seniority-zero space.
	\begin{subequations}
		\begin{eqnarray}
			K^{\Pi}_{ab} &=& \frac{2}{N-1} h_{ab}\delta_{ab} + V_{aabb} \\
			K^D_{ab} &=& \frac{2}{N-1} (h_{aa} + h_{bb}) + 2V_{abab} - V_{abba}
		\end{eqnarray}
	\end{subequations}
	
	\section{$N$-representability optimization algorithm} \label{sec:algorithm}
	For wave function methods that are solved in a projective manner, like CC and geminals methods, the approximate 2-RDMs obtained from response theory, Eq.~(\ref{resp2dm}), are in general not $N$-representable. Consequently, they do not necessarily satisfy the $N$-representability constraints presented in the previous section, nor do they always have the right symmetry. The goal of the new algorithm we will present here is to ``fix'' the response 2-RDM so it becomes approximately $N$-representable, \textit{i.e.} it has the right symmetry, Eqs.~(\ref{sym2rdm}) and (\ref{antisym2rdm}), the right trace, Eq.~(\ref{trace2rdm}), and it satisfies the $P$-, $Q$- and $G$-conditions, Eqs.~(\ref{Pcond}), (\ref{Qcond}) and (\ref{Gcond}). This optimization problem can actually be formulated as a semidefinite programming problem,
	\begin{equation} 
		\label{sdp}
		\begin{aligned}
			&\underset{\Gamma}{\text{minimize}}
			& & \|\Gamma-\Gamma^{response} \|^2 \\
			&\text{subject to} & &\Gamma = \Gamma^T \\
			&&&\mathrm{Tr}(\Gamma) = N(N-1) \\
			&&& \Gamma, Q(\Gamma), G(\Gamma) \, \succeq \, 0
		\end{aligned}
	\end{equation}
	Our strategy is to find the 2-RDM that minimizes the square norm of the difference between the initial 2-RDM and the targeted 2-RDM, under the constraints that this 2-RDM is symmetric, its trace is normalized and the $P$, $Q$- and $G$-matrices are positive semidefinite matrices, \textit{i.e.} their eigenvalues are non-negative. This kind of problem strongly resembles the variational optimization of the energy in 2-RDM theory. \cite{mazziotti2007advances,verstichel2009v2dm,verstichel2011v2dm}
	
	Given the problem definition in Eq.~(\ref{sdp}) we could approach this in the context of semidefinite programming (SDP).\cite{nakata2001v2dm,nakata2002v2dm,vandenberghe1996sdp, boyd2004sdp} However, the methods for solving SDP problems are quite complex and existing general purpose solvers are inefficient for large-scale problems. Given that the initial 2-RDM comes from a computationally efficient method (\textit{e.g.} AP1roG), we prefer a``fixing'' procedure of comparable scaling. So, in this section we propose an alternative approach to find an approximate solution to Eq.~(\ref{sdp}) which is computationally much faster. 
		
	In 1988, Higham\cite{higham1988sdp} showed that for a given real matrix $B_0$, the closest (with respect to the Frobenius norm) symmetric positive semidefinite matrix $B$, \textit{i.e.} the solution to
	\begin{equation} \label{higham}
	\begin{aligned}
	&\underset{B}{\text{minimize}}
	& & \| B - B_0\|^2 \\
	& \text{subject to} & &  B = B^T \\
	&&& B \, \succeq \, 0
	\end{aligned}
	\end{equation}
	is given by the positive symmetric part of $B_0$. However, Higham's method does not preserve the normalization of the initial matrix $B_0$. From the problem definition in Eq.~(\ref{sdp}) we see that we need to impose an additional constraint on the trace, \textit{i.e.} we are dealing with a problem of the following kind:
	\begin{equation} \label{highamext}
	\begin{aligned}
	&\underset{B}{\text{minimize}}
	& & \| B - B_0\|^2 \\
	& \text{subject to} & & B = B^T \\
	&&& B \, \succeq \, 0 \\
	&&& \mathrm{Tr}B = T
	\end{aligned}
	\end{equation}
	with $T$ a known fixed value. We can show that the closest symmetric positive semidefinite matrix $B$ with fixed trace is obtained by shifting the set of eigenvalues $\{\lambda_i\}$ of $B_0$ with a value $\sigma_0$ which is the root of the equation $f(\sigma) = T$ where $f(\sigma) = \sum_i\theta(\lambda_i - \sigma) (\lambda_i - \sigma)$ with $\theta(\lambda_i - \sigma)$ the Heaviside step function. A proof for both Higham's problem (Eq.~(\ref{higham})) and the optimization problem from Eq.~(\ref{highamext}) with the additional constraint on the trace, is given in the Appendix.
	
	Based on this result, we propose the following iterative procedure to determine the closest, positive semidefinite, symmetric 2-RDM with the correct trace:
	\begin{enumerate}[label=\emph{\alph*})]
		\item Symmetrize the 2-RDM:
		\begin{equation} \label{2pbsym}
		\Gamma^{response}_{sym} = \frac{1}{2}\left( \Gamma^{response} + \left(\Gamma^{response}\right)^{T}\right)
		\end{equation}
		\item Compute the eigenvector decomposition to find the eigenvalues $\{\lambda\}$:
		\begin{equation}
		\Gamma^{response}_{sym} = UDU^T
		\end{equation}
		\item Shift all the eigenvalues by a constant $\sigma_0$ which is the root of the equation $f(\sigma) = \sum_i \theta(\lambda_i - \sigma)(\lambda_i - \sigma) = \mathrm{Tr}(\Gamma)$ and set any negative shifted eigenvalues to zero.
		\begin{eqnarray}
		\lambda^+ =
			\begin{cases}
				\lambda - \sigma_0 \quad &\text{if } \lambda > \sigma_0 \\
				0 \quad &\text{if } \lambda \leq \sigma_0
			\end{cases}
		\end{eqnarray}
		The root $\sigma_0$ is determined through a bisection method.
		\item Find an updated 2-RDM from the resulting shifted set of eigenvalues and the original eigenvectors
		\begin{equation}
		\tilde{\Gamma} = U\tilde{D}U^{T}
		\end{equation}
	\end{enumerate}
	In order for the 2-RDM to also satisfy the $Q$- and $G$-condition, a similar procedure can be used for making the $Q$- and $G$-matrix positive semidefinite under the normalization constraints (\ref{traceq}) and (\ref{traceg}).  Alternatively, instead of determining the shift for the eigenvalues through a bisection method, we can also determine the shifted eigenvalues in an iterative fashion: in a first step we set the negative eigenvalues to zero and in a second step all the eigenvalues are shifted by a constant to fulfil the trace conditions. These two steps are repeated until none of the eigenvalues are negative anymore and we have the correct trace.
	
\begin{figure}
		\includegraphics{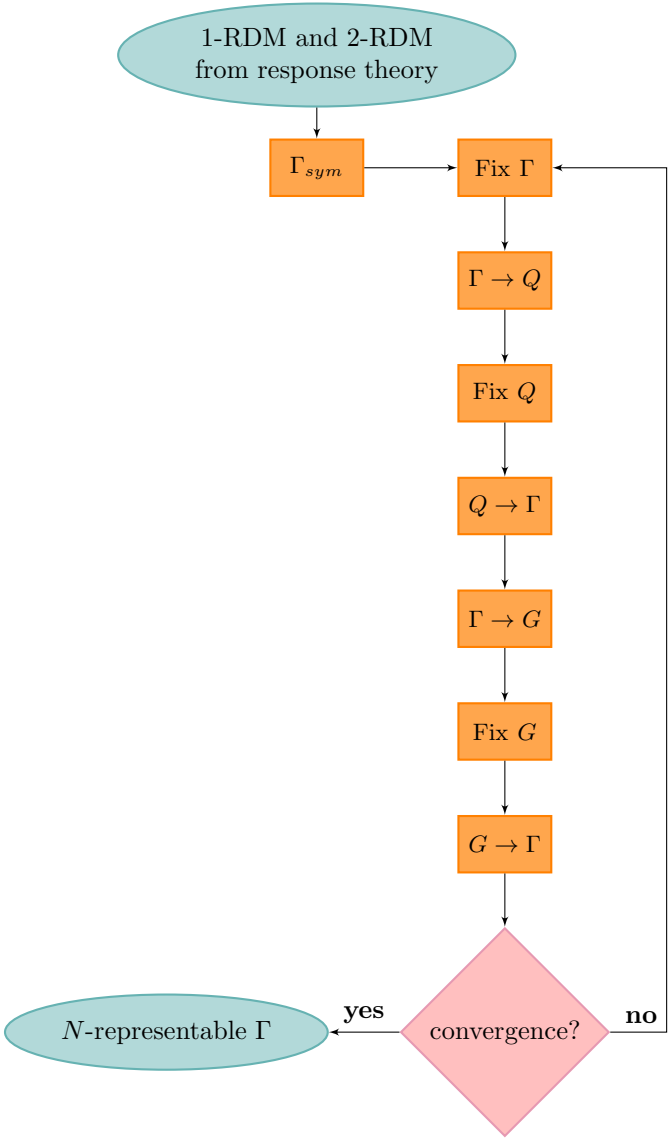}
		\caption{\label{fig:general_algorithm} General scheme for the $N$-representability optimization procedure.}
	\end{figure}

	An overview of the $N$-representability optimization algorithm is given in Fig.~\ref{fig:general_algorithm}; it is referred to as the regular 2-RDM algorithm. We sequentially optimize the $P$-, $Q$- and $G$-matrix and iterate until the procedure converges. Convergence is measured by the error on the traces and the magnitude of the largest negative eigenvalue. When convergence is reached we have determined the 2-RDM which is close to the initial 2-RDM, but which is $N$-representable in the sense that it fulfills the $P$-, $Q$- and $G$-conditions. We will refer to this converged 2-RDM as the ``fixed'' 2-RDM. We emphasize that the obtained 2-RDM from the optimization algorithm is not strictly the closest in the sense of Eq.~(\ref{sdp}), but corresponds to a 2-RDM that is also close, has the right $P$-, $Q$- and $G$-properties, and is obtained at a much more favourable computational cost compared to the exact solution, Eq.~(\ref{sdp}). It is possible to change the order in which the $P$-, $Q$- and $G$-matrices are fixed, however, we observed that this does not significantly influence the results.
	
	In the case of seniority-zero space, the non-zero elements of the 2-RDM are split over the pair density matrix $\Pi$ and $D$ and the optimization problem can be formulated as
	\begin{equation} 
		\label{sdp_doci}
		\begin{aligned}
			&\underset{\Pi, D}{\text{minimize}}
			& & \left(\|\Pi-\Pi^{response} \|^2 + \|D - D^{response}\|^2\right) \\
			&\text{subject to} & &\Pi = \Pi^T , \quad D = D^T \\
			&&&\mathrm{Tr}(\Pi) = \frac{N}{2}, \quad \mathrm{Tr}(D) = \frac{N}{2}(\frac{N}{2} - 1) \\
			&&& \Pi, Q^{\Pi}, G^{\Pi}, G^{(2x2)} \, \succeq \, 0 \\
			&&& D_{ab}, Q^D_{ab} \geq 0 \quad (\forall a \neq b)
		\end{aligned}
	\end{equation}
	We try to fix both the pair and exchange density matrix, $\Pi$ and $D$, such that they become approximately $N$-representable in the sense that they are symmetric, they have the right normalization, Eqs.~(\ref{tracepair2rdm}) and (\ref{traceexch2rdm}), and they satisfy the DOCI-reformulated $P$-, $Q$-, and $G$-conditions presented in section \ref{sec:dociNrepconditions}. This algorithm is denoted as the DOCI 2-RDM algorithm. As in the regular 2-RDM optimization algorithm, the $P$-, $Q$- and $G$-condition are enforced sequentially. Convergence is measured by the norm of the difference between density matrices from consecutive iterations.
	
	\section{Results} \label{sec:results}
	The algorithms for making the 2-RDM approximately $N$-representable are tested on a variety of systems. We start with applying the DOCI 2-RDM algorithm on a typical example for strong correlation: the symmetric dissociation of a linear chain of equally spaced hydrogen atoms. We then look at some 12- and 14-electron diatomic species where we perform calculations for three different bond lengths with both the DOCI 2-RDM (doci2RDM) and regular 2-RDM (reg2RDM) algorithm. Finally, we pushed the DOCI 2-RDM optimization algorithm into a regime where AP1roG is known to fail, \textit{i.e.} the attractive Richardson pairing Hamiltonian.\cite{richardson1963pairinghamI,richardson1964pairinghamII}
	The Hartree-Fock and AP1roG calculations are done with the HORTON software package.\cite{horton} For the AP1roG calculations, a variational orbital optimization scheme,\cite{boguslawski2014geminals,boguslawski2014ap1rog,boguslawski2014nonvariational} is used. The full CI energies are obtained with the Psi4 program package.\cite{psi4} All DOCI calculations are done with in-house developed programs. 
	
	For comparing the fixed 2-RDM, $\tilde{\Gamma}$, with the initial 2-RDM, $\Gamma^{response}$, we define a cost function, weighted by the number of terms in the summation
	\begin{equation} \label{costfunc}
	f = \frac{1}{(2L)^4}\sum_{ijkl} \left(\Gamma_{ijkl}^{response} - \tilde{\Gamma}_{ijkl}\right)^2
	\end{equation}
	and in the case of DOCI space
	\begin{equation} \label{costfunc_doci}
	f = \frac{1}{2L^2}\sum_{ij} \left[\left(\Pi_{ij}^{response} - \tilde{\Pi}_{ij}\right)^2 + \left(D_{ij}^{response} - \tilde{D}_{ij}\right)^2\right]
	\end{equation}
	\subsection{Linear hydrogen chain, $\text{H}_8$}
\begin{figure}
		\includegraphics[scale=0.5]{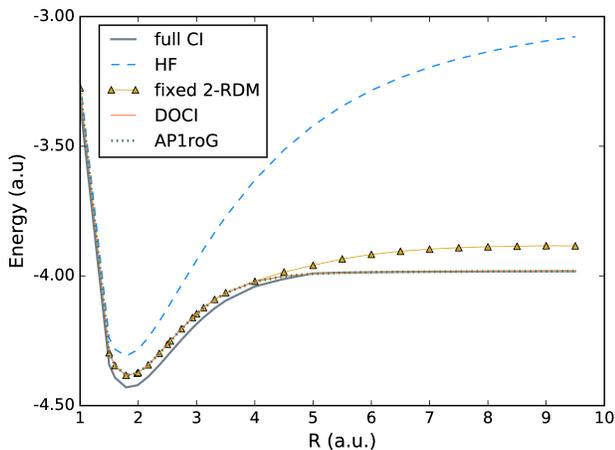}
		\caption{\label{fig:h8_diss} Symmetric dissociation of a linear $\text{H}_8$ chain with interatomic distance $R$ using an ANO-2s basis set.\cite{widmark1990ano2s} Total energies predicted by full CI, Hartree-Fock, AP1roG (with optimized orbitals), DOCI (with optimized orbitals) and fixed 2RDM energy (calculated according to Eq.~(\ref{doci_energy})).}
	\end{figure}

	In Fig.~\ref{fig:h8_diss}, the energy profile for the dissociation of a linear $\text{H}_8$ chain is shown. The AP1roG and DOCI results are nearly indistinguishable over the whole range of interatomic distances. Both are also very close to the full CI solution; the deviation is due to the lack of dynamical correlation in these seniority-zero methods.\cite{bytautas2011seniority,limacher2013geminals} Subsequently, we use the DOCI 2-RDM algorithm to fix the AP1roG response 2-RDM and the energy is calculated according to Eq.~(\ref{doci_energy}). Note that using directly the response 2-RDM in Eq.~(\ref{doci_energy}) reproduces the AP1roG energy and is therefore not plotted separately.
	
	For interatomic distances up to 4 a.u. the fixed 2-RDM energy, AP1roG energy and DOCI energy are within less than milliHartree (mHa) range. For larger bond distances, the fixed 2-RDM energy starts to deviate from the AP1roG and DOCI energy and the difference increases to 0.1 Ha. During the optimization procedure, the 2-RDM is restricted to fulfill some conditions, hence energy increases are to be expected. The small deviations around equilibrium length indicate that the original 2-RDMs are in fact very close to being $N$-representable, while at larger bond lengths the response 2-RDMs are further away from $N$-representability. As a measure for the violation of the 2-positivity $N$-representability conditions of the initial 2-RDM, we can use the (absolute) sum of all negative eigenvalues of the $P$-, $Q$- and $G$-matrix. This sum is below $6 \times 10^{-4}$ for bond distances up to 4 a.u. and for larger bond lengths it varies from 0.15 to almost 1.00. The larger the interatomic distances, the more the response 2-RDM violates the $P$-, $Q$- and $G$-conditions. Although the AP1roG energy is an excellent approximation for the DOCI energy, also in the bond breaking region, the structure of the response 2-RDM deviates more and more from $N$-representability with increasing bond length. This is also confirmed by the cost functions. For bond lengths under 4 a.u. the cost function was found to be below $5 \times 10^{-9}$ and at larger bond lengths it increases from $10^{-5}$ up to values around $2.6\times10^{-3}$ thereby confirming that at larger bond lengths the fixed, $N$-representable 2-RDMs differ more from the initial 2-RDMs than around the equilibrium distance. 
	
	\subsection{12- and 14-electron species}
		The DOCI 2-RDM algorithm is tested on seven 12- and 14-electron molecules. Table~\ref{tab:12/14species_doci2rdm} lists the DOCI energy (with AP1roG optimized orbitals) and the energy difference with respect to the AP1roG energy, $\Delta E_{AP1roG}$, for three interatomic distances. It also shows the difference of the fixed DOCI 2-RDM energy and AP1roG energy, $\Delta E_{doci2RDM}$. As was already shown in earlier papers,\cite{limacher2013geminals,tecmer2014accuracy,tecmer2015actinide} the AP1roG energy is a very good approximation to the DOCI energy, it deviates at most by about 4.9 mHa from the DOCI energy. Finally, the cost functions, Eq.~(\ref{costfunc_doci}), are also listed in Table~\ref{tab:12/14species_doci2rdm}. For equilibrium geometries, the response 2-RDM is very close to being $N$-representable. For larger bond lengths, the cost functions increase, indicating larger corrections are needed in order to enforce $N$-representability of the initial 2-RDM. 
	
	\begin{table*}
		\centering
		\caption{DOCI energy (with AP1roG optimized orbitals), energy difference with respect to AP1roG, $\Delta E_{AP1roG} = E_{DOCI} - E_{AP1roG}$, energy difference of the AP1roG energy and the 2-RDM fixed through the DOCI 2-RDM algorithm, $\Delta E_{doci2RDM} = E_{doci2RDM} - E_{AP1roG}$ and the cost function as defined in Eq.~(\ref{costfunc_doci}). All calculations are done using the 6-31g* basis set and all energies are given in atomic units. The equilibrium bond lengths, $R_e$, are taken from the NIST CCCBDB\cite{cccbdb} database for CCSD(T) optimized geometries.}
		\label{tab:12/14species_doci2rdm}
		\begin{tabular}{llllllll}
			\hline
			\multicolumn{1}{c}{molecule}    & \multicolumn{1}{c}{$\text{C}_2$}  & \multicolumn{1}{c}{BN}     & \multicolumn{1}{c}{BeO}    & \multicolumn{1}{c}{LiF}    & \multicolumn{1}{c}{$\text{N}_2$}  & \multicolumn{1}{c}{CO}     & \multicolumn{1}{c}{BF}     \\
			\multicolumn{1}{c}{$R_e$(\AA)} & \multicolumn{1}{c}{1.2562} & \multicolumn{1}{c}{1.3369} & \multicolumn{1}{c}{1.3490} & \multicolumn{1}{c}{1.5658} & \multicolumn{1}{c}{1.1191} & \multicolumn{1}{c}{1.1472} & \multicolumn{1}{c}{1.2829} \\ \hline
			& \multicolumn{7}{c}{Interatomic distance $R = R_e$} \\ 
			$E_{DOCI}$				& $-75.5416$			& $-79.0162$			& $-89.4948$			& $-107.0007$			& $-109.0597$			& $-112.8392$			& $-124.2000$ \\
			$\Delta E_{AP1roG}$		& $3.0\times10^{-5}$	& $2.8\times10^{-4}$	& $2.0\times10^{-5}$	& $-6.3\times10^{-6}$	& $-8.4\times10^{-5}$	& $-2.9\times10^{-5}$	& $-1.7\times10^{-5}$ \\
			$\Delta E_{doci2RDM}$	& $8.5\times10^{-3}$	& $7.4\times10^{-3}$	& $4.4\times10^{-3}$	& $6.4\times10^{-3}$	& $9.5\times10^{-3}$	& $8.5\times10^{-3}$	& $2.6\times10^{-4}$ \\
			$f_{doci2RDM}$ 			& $4.7\times10^{-7}$	& $1.7\times10^{-7}$	& $4.9\times10^{-10}$ 	& $1.0\times10^{-10}$ 	& $1.2\times10^{-9}$ 	& $5.0\times10^{-10}$ 	& $1.7\times10^{-11}$ \\ \hline
			& \multicolumn{7}{c}{Interatomic distance $R = R_e + 1.00$\AA} \\
			$E_{DOCI}$				& -75.3490				& -78.8974				& -89.3491				& -106.9057				& -108.7107				& -112.4785				& -124.0251 \\
			$\Delta E_{AP1roG}$		& $8.1\times10^{-5}$	& $6.8\times10^{-5}$	& $-1.6\times10^{-4}$	& $ 2.0\times10^{-6}$	& $ 1.5\times10^{-3}$	& $4.9\times10^{-3}$	& $1.1\times10^{-4}$ \\
			$\Delta E_{doci2RDM}$	& $2.1\times10^{-1}$	& $1.6\times10^{-1}$	& $5.0\times10^{-2}$	& $7.3\times10^{-3}$	& $4.2\times10^{-1}$	& $2.9\times10^{-1}$	& $5.8\times10^{-2}$ \\
			$f_{doci2RDM}$			& $5.7\times10^{-6}$ 	& $7.8\times10^{-6}$ 	& $5.2\times10^{-8}$	& $2.4\times10^{-10}$	& $9.8\times10^{-6}$	& $3.4\times10^{-6}$	& $5.4\times10^{-8}$ \\ \hline
			& \multicolumn{7}{c}{Interatomic distance $R=5.00$\AA} \\
			$E_{DOCI}$ 				& -75.3051				& -78.8536				& -89.3477				& -106.8345				& -108.5920				& -112.4708				& -123.9567 \\ 
			$\Delta E_{AP1roG}$		& $-6.2\times10^{-4}$	& $1.1\times10^{-5}$ 	& $-3.8\times10^{-5}$	& $2.3\times10^{-5}$	& $ 2.6\times10^{-4}$ 	& $1.2\times10^{-3}$ 	& $2.1\times10^{-5}$ \\
			$\Delta E_{doci2RDM}$	& $8.9\times10^{-1}$ 	& $4.1\times10^{-1}$ 	& $1.3\times10^{-1}$ 	& $1.2\times10^{-3}$ 	& $2.5\times10^{0}$ 	& $9.1\times10^{-1}$ 	& $3.2\times10^{-2}$ \\
			$f_{doci2RDM}$			& $1.0\times10^{-4}$	& $2.9\times10^{-5}$	& $6.0\times10^{-6}$	& $5.9\times10^{-9}$	& $5.6\times10^{-4}$	& $4.1\times10^{-5}$	& $1.8\times10^{-7}$ \\  \hline
		\end{tabular}
	\end{table*}
		
		Since the AP1roG energy is such a good approximation to the DOCI energy, it would be preferable to have only slight changes in energy upon fixing the 2-RDM. In line with the results from the dissociation of a linear $\text{H}_8$ chain, there is an increase in energy, compared to AP1roG, as a result from constraining the 2-RDM during the $N$-representability optimization procedure. The reported energy difference, $\Delta E_{doci2RDM}$, can be seen as an energetic cost for enforcing $N$-representability on the initial 2-RDM. This energetic cost is of the order milliHartree for equilibrium bond lengths and increases with increasing interatomic distance. It even attains the unrealistic value of 2.5 Hartree for the stretched $\text{N}_2$ molecule. Moreover, it appears the method performs worse for stretched diatomic species with multiple bonds, like $\text{N}_2$ and CO, while its performance is better for singly bonded diatomics, like LiF and BF.
	
		By using the DOCI 2-RDM algorithm, we restrict the 2-RDM to seniority-zero space, but this leads to large increases in energy, mostly for stretched geometries. Therefore, we have also tested the regular 2-RDM fixing procedure on these seven diatomic molecules such that the 2-RDM can move outside seniority-zero space. We should note that for this algorithm the full 2-RDM, which is a four-dimensional array of size $2L$, is constructed. As such, the memory requirements are much higher compared to the DOCI specific algorithm where we are just required to store two $L\times L$ matrices. The results can be found in Table~\ref{tab:12/14species_reg2rdm}. The cost functions, Eq.~(\ref{costfunc}), are much lower in comparison to the DOCI 2-RDM algorithm. We find an $N$-representable 2-RDM which is closer to the initial 2-RDM compared to the fixed DOCI 2-RDM. However, this fixed reg2-RDM still has a DOCI structure (see Table~\ref{tab:12/14species_sum}) meaning that it still lies in the seniority-zero space. The regular 2-RDM fixing procedure converges to a 2-RDM which is also in DOCI space, however, it is a different solution than the one which results from the DOCI 2-RDM algorithm. The energetic cost due to the regular 2-RDM fixing procedure is lower than the energy increase due to DOCI 2-RDM fixing. However, it is still quite high, particularly for the stretched geometries, \textit{i.e.} about 0.9 Hartree for $\text{N}_2$. 
	
	\begin{table*}
		\centering
		\caption{Energy difference of the AP1roG energy and the 2-RDM fixed through the regular 2-RDM algorithm, $\Delta E_{reg2RDM} = E_{reg2RDM} - E_{AP1roG}$ and the cost function as defined in Eq.~(\ref{costfunc}).}
		\label{tab:12/14species_reg2rdm}
		\begin{tabular}{llllllll}
			\hline
			\multicolumn{1}{c}{molecule}    & \multicolumn{1}{c}{$\text{C}_2$}  & \multicolumn{1}{c}{BN}     & \multicolumn{1}{c}{BeO}    & \multicolumn{1}{c}{LiF}    & \multicolumn{1}{c}{$\text{N}_2$}  & \multicolumn{1}{c}{CO}     & \multicolumn{1}{c}{BF}     \\
			\multicolumn{1}{c}{$R_e$(\AA)} & \multicolumn{1}{c}{1.2562} & \multicolumn{1}{c}{1.3369} & \multicolumn{1}{c}{1.3490} & \multicolumn{1}{c}{1.5658} & \multicolumn{1}{c}{1.1191} & \multicolumn{1}{c}{1.1472} & \multicolumn{1}{c}{1.2829} \\ \hline
			& \multicolumn{7}{c}{Interatomic distance $R = R_e$} \\ 
			$\Delta E_{reg2RDM}$	& $1.5\times10^{-3}$	& $8.3\times10^{-3}$	& $6.9\times10^{-5}$	& $1.4\times10^{-5}$	& $2.4\times10^{-4}$	& $1.2\times10^{-4}$	& $4.0\times10^{-5}$ \\
			$f_{reg2RDM}$			& $8.9\times10^{-13}$	& $3.9\times10^{-13}$	& $6.3\times10^{-16}$ 	& $3.0\times10^{-17}$ 	& $2.9\times10^{-15}$ 	& $1.1\times10^{-15}$ 	& $2.5\times10^{-16}$ \\ \hline
			& \multicolumn{7}{c}{Interatomic distance $R = R_e + 1.00$\AA} \\
			$\Delta E_{reg2RDM}$	& $2.2\times10^{-2}$	& $7.5\times10^{-3}$	& $5.9\times10^{-4}$	& $2.2\times10^{-5}$	& $4.2\times10^{-2}$	& $2.1\times10^{-2}$	& $5.4\times10^{-4}$ \\
			$f_{reg2RDM}$			& $1.2\times10^{-10}$ 	& $1.5\times10^{-11}$ 	& $1.9\times10^{-14}$	& $7.4\times10^{-17} $	& $4.4\times10^{-10}$	& $1.2\times10^{-10}$	& $1.0\times10^{-13}$ \\ \hline
			& \multicolumn{7}{c}{Interatomic distance $R=5.00$\AA} \\
			$\Delta E_{reg2RDM}$	& $9.2\times10^{-2}$ 	& $1.4\times10^{-1}$ 	& $5.6\times10^{-3}$ 	& $9.5\times10^{-5}$ 	& $9.1\times10^{-1}$ 	& $1.3\times10^{-1}$ 	& $5.0\times10^{-4}$ \\
			$f_{reg2RDM}$			& $4.5\times10^{-9}$	& $1.8\times10^{-8}$	& $9.3\times10^{-12}$	& $2.6\times10^{-15}$	& $6.3\times10^{-7}$	& $1.4\times10^{-8}$	& $3.2\times10^{-14}$ \\ \hline
		\end{tabular}
	\end{table*} 
	
	\begin{table*}
		\centering
		\caption{Sum of the elements of the final 2-RDM fixed through the regular 2-RDM algorithm that should be zero in case of DOCI structure, \textit{i.e.} elements that are not of the type in Eqs.~(\ref{doci_pair}) and (\ref{doci_exchange}).}
		\label{tab:12/14species_sum}
		\begin{tabular}{lrrrrrrr}
			\hline
			\multicolumn{1}{c}{molecule}    & \multicolumn{1}{c}{$\text{C}_2$}  & \multicolumn{1}{c}{BN}     & \multicolumn{1}{c}{BeO}    & \multicolumn{1}{c}{LiF}    & \multicolumn{1}{c}{$\text{N}_2$}  & \multicolumn{1}{c}{CO}     & \multicolumn{1}{c}{BF}     \\
			\multicolumn{1}{c}{$R_e$(\AA)} & \multicolumn{1}{c}{1.2562} & \multicolumn{1}{c}{1.3369} & \multicolumn{1}{c}{1.3490} & \multicolumn{1}{c}{1.5658} & \multicolumn{1}{c}{1.1191} & \multicolumn{1}{c}{1.1472} & \multicolumn{1}{c}{1.2829} \\ \hline 
			$R = R_e$			& $ 6.6\times10^{-9}$	& $ 6.3\times10^{-9}$	& $ 1.1\times10^{-8}$	& $ 5.7\times10^{-9}$	& $ 1.7\times10^{-8}$	& $ 1.5\times10^{-8}$	& $ 1.4\times10^{-8}$ \\
			$R = R_e + 1.00$\AA	& $ 2.1\times10^{-8}$	& $ 8.7\times10^{-9}$	& $ 1.8\times10^{-8}$	& $ 8.5\times10^{-9}$	& $ 9.6\times10^{-9}$	& $ 8.4\times10^{-9}$	& $ 1.8\times10^{-8}$ \\
			$R = 5.00$\AA		& $ 1.3\times10^{-8}$	& $ 1.5\times10^{-8}$	& $ 9.4\times10^{-9}$	& $ 1.3\times10^{-8}$	& $ 1.4\times10^{-8}$	& $ 9.0\times10^{-9}$	& $ 6.0\times10^{-9}$ \\ \hline
		\end{tabular}
	\end{table*}

\begin{figure}
		\includegraphics[scale=0.5]{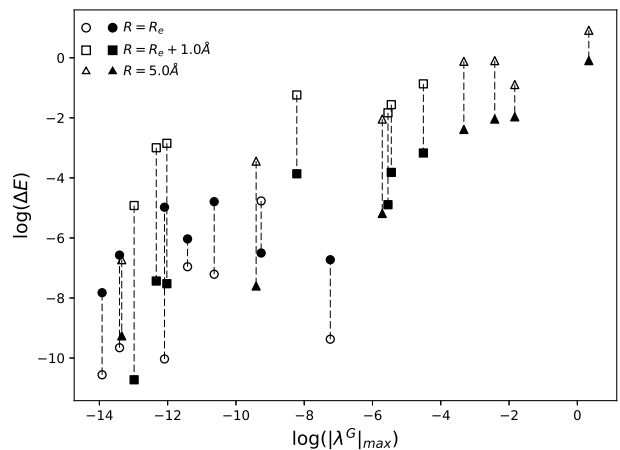}
		\caption{\label{fig:12/14species} Log-log scatter plot of the absolute value of the most negative eigenvalue of the initial $G$-matrix and the $\Delta E_{doci2RDM}$ (unfilled markers) and the $\Delta E_{reg2RDM}$ (filled markers) at three interatomic distances.}
	\end{figure}

	We found that the deviation from $N$-representability of the initial 2-RDM is mostly due to violation of the $G$-condition. In Fig.~\ref{fig:12/14species} a log-log scatter plot is shown of the most negative eigenvalue of the initial $G$-matrix and the energetic cost due to DOCI 2-RDM fixing and the regular 2-RDM fixing. The violation of the G-condition is strongest for stretched molecules, and this can be correlated with a higher energy increase upon 2-RDM fixing. As could already be concluded, the use of the regular 2-RDM algorithm reduces the energy increase compared to the DOCI 2-RDM algorithm, especially at larger interatomic distances.
	
	\subsection{Pairing Hamiltonian}
	Until now we have handled systems where the AP1roG model is known to perform well for energies. In general, the resulting 2-RDMs did not deviate too much from $N$-representability. It has been shown that AP1roG fails for the attractive pairing model as soon as a critical pairing interaction strength is reached.\cite{dukelsky2003pairingham, henderson2014pairingham} The Richardson pairing Hamiltonian can be written as\cite{richardson1963pairinghamI,richardson1964pairinghamII}
	\begin{equation}
	\hat{H} = \sum_p \epsilon_p \padd{p}\hat{a}_{p} + g \sum_{pq}S_p^{\dagger}S_q
	\end{equation}
	The parameter $g$ is called the interaction strength and can distinguish between attractive ($g < 0$) or repulsive ($g > 0$) interactions. The set $\{ \epsilon_i \}$ are the single particle energies. Specifically, we will investigate the half-filled pairing Hamiltonian with twelve equally spaced levels, the so-called picket-fence model,\cite{sambataro2007picketfence,stijn2012RGpicketfence} \textit{i.e.} $\epsilon_p = p \, \Delta \epsilon$ with $\Delta \epsilon$ the interlevel spacing. In this case the ground state lies in the fully paired (seniority-zero) space, thus DOCI coincides with full CI. For the following calculations we again use the DOCI 2-RDM algorithm. In this subsection, the AP1roG and full CI energies are calculated with an in-house program.
	
\begin{figure}
		\includegraphics[scale=0.5]{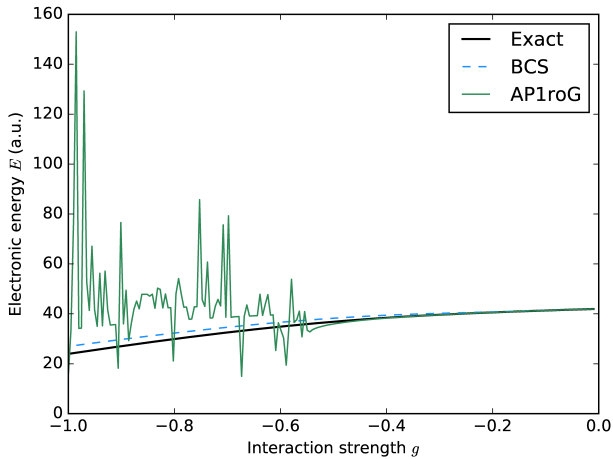}
		\caption{\label{fig:pairing_attractive_energy} Energy as a function of the interaction strength $g$ for the attractive pairing model ($g < 0$) of six electron pairs in twelve orbitals with equally spaced single particle energies. Total energies predicted by the AP1roG and BCS methods and the exact solution according to Richardson.}
	\end{figure}

	In Fig.~\ref{fig:pairing_attractive_energy}, the total energy as a function of the interaction strength $g$ in the attractive region is shown. It is clear that beyond a certain interaction strength $g$ AP1roG completely fails to correctly describe the system. This is because the particle-hole reference states are not sufficient to correctly describe the collective pairing oscillations of the superconducting ground state of the system. However, the plotted energies are still converged solutions to the AP1roG equations and may correspond to non-collective excited states of the system. Next to the exact solution according to Richardson,\cite{richardson1963pairinghamI, richardson1964pairinghamII} we also show the Bardeen-Cooper-Schrieffer (BCS) solution.\cite{bardeen1957bcs} AP1roG breaks down near a critical pairing interaction strength of $g_c \cong -0.35$ which is the transition point to a superconducting system \textit{i.e.} the HF to BCS transition point where the superconducting gap is of the same order as the interlevel spacing.\cite{sambataro2007picketfence,stijn2012RGpicketfence} 
	
\begin{figure}
		\includegraphics[scale=0.5]{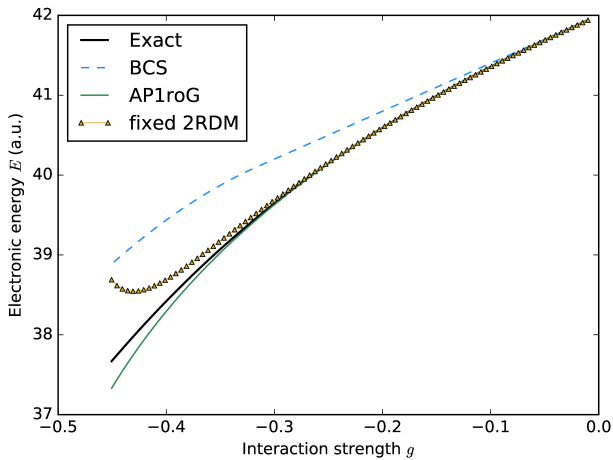}
		\caption{\label{fig:pairing_attractive_energy_zoom} Same data as in Fig.~\ref{fig:pairing_attractive_energy} with the additional curve representing the fixed 2-RDM energy.}
	\end{figure}

\begin{figure}
		\includegraphics[scale=0.5]{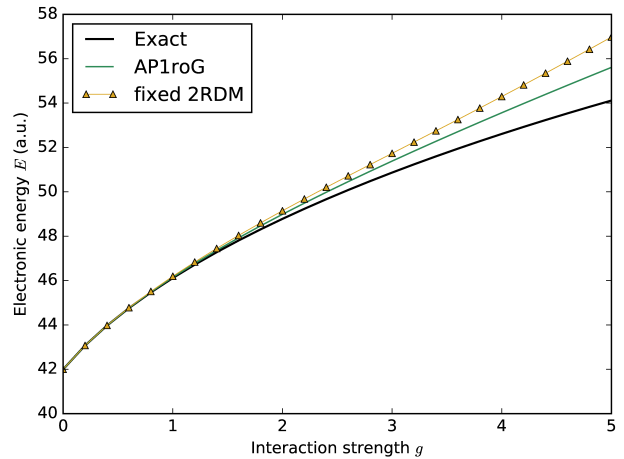}
		\caption{\label{fig:pairing_repulsive_energy} Energy as a function of the interaction strength $g$ for the repulsive pairing model ($g > 0$) for the same system as in Fig.~\ref{fig:pairing_attractive_energy}. Total energies from AP1roG method, fixed 2-RDM and the exact solution according to Richardson.}
	\end{figure}

	In Fig.~\ref{fig:pairing_attractive_energy_zoom}, one sees that when AP1roG starts to fail, the energy even goes below the exact solution. This is indeed possible since AP1roG is not a variational method, for which the exact energy is an upper limit, but is solved through the projected Schr\"{o}dinger equation. Fixing the 2-RDM partly remedies the non-variational character of the AP1roG energy, however, the increase in energy becomes more pronounced reaching the critical $g_c$ signalling the inadequacy of AP1roG in this regime. $N$-representability fixing of the 2-RDM results in energies that are above the exact energy.
	We also show the energy profiles for the repulsive pairing model in Fig.~\ref{fig:pairing_repulsive_energy}, $g > 0$. Here AP1roG does not fail as drastically as for the attractive region because the ground state is better characterized by a few particle-hole excitations across the reference state. $N$-representability fixing of the 2-RDM again increases the energy, which is in line with the results obtained for the previous systems.
	
\begin{figure}
		\includegraphics[scale=0.5]{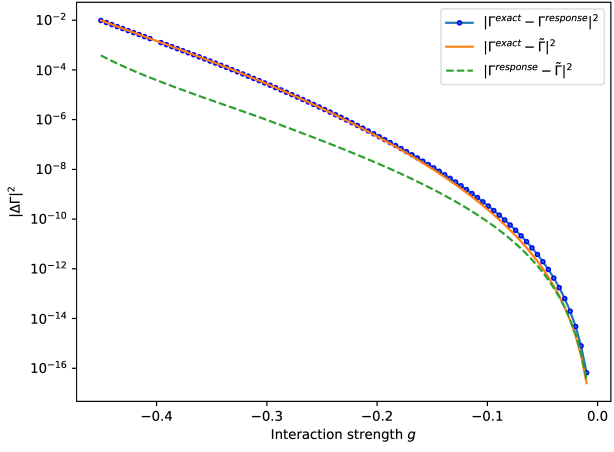}
		\caption{\label{fig:costfunc_attractive} Cost functions of the AP1roG response 2-RDM and the fixed 2-RDM with respect to the exact 2-RDM and of the response 2-RDM with respect to the fixed 2-RDM as a function of the interaction strength $g$ for the attractive pairing model.}
	\end{figure}

	For the integrable pairing Hamiltonian we are able to compute the exact 2-RDMs at fairly low cost, so we can now compare the initial 2-RDM and fixed 2-RDM with the exact 2-RDM, see Fig.~\ref{fig:costfunc_attractive}. The cost functions are defined in a similar fashion as in Eq.~(\ref{costfunc_doci}), but now with respect to the exact 2-RDM. We also plot the cost function of the response 2-RDM and fixed 2-RDM. As expected the cost functions increase with increasing (in absolute value) interaction strength. Moreover, the cost functions of the response 2-RDM and fixed 2-RDM with respect to the exact 2-RDM nearly overlap indicating that although we bring the 2-RDM closer to being $N$-representable, the distance to the exact 2-RDM does not change much. This could indicate that the $N$-representability optimization algorithm seems to bring the 2-RDM closer to the set of $N$-representable 2-RDMs but not necessarily closer to the exact 2-RDM.  

	\section{Conclusions}
	Many methods in quantum chemistry rely on a projective method to solve the Schr\"odinger equation, like CC and geminal-based methods. Since for these methods, direct computation of the 2-RDM is impractical, it may be better to compute the 2-RDM from response theory. However, these approximate 2-RDMs often suffer from the so-called $N$-representability problem. In this manuscript we have introduced an optimization algorithm to make non-$N$-representable 2-RDMs approximately $N$-representable. Our algorithm presents an alternative approach to semidefinite programming problems. Although it only provides an approximate solution, it is computationally less complex and much faster. Due to the current interest in seniority-zero wave function models, we also developed a 2-RDM optimization procedure tailored specifically to DOCI space. With the proposed algorithms we can find the 2-RDM which is closest (w.r.t. the Frobenius norm) to the  initial non-$N$-representable 2-RDM under the constraints that the resulting 2-RDM should be symmetric, its trace has the correct normalization and it satisfies the standard 2-positivity conditions.
	
	From the results we can conclude that for bond breaking processes, where the AP1roG energy is a very good approximation for the DOCI energy, the structure of the response 2-RDM deviates more and more from $N$-representability with increasing interatomic distances. Fixing the 2-RDMs results in an energy increase with respect to the original 2-RDM energy because we put certain constraints on the 2-RDM during the $N$-representability optimization procedure. Using the general 2-RDM algorithm to enforce $N$-representability has a lower energetic cost compared to the DOCI 2-RDM algorithm, however for stretched geometries this energy increase is quite high for both algorithms. For the attractive Richardson pairing Hamiltonian, AP1roG fails beyond a certain critical interaction strength resulting in response 2-RDMs which are far from being $N$-representable. We are able to compare both the response 2-RDM and fixed 2-RDM with the exact solution and it seems that although we have made the initial 2-RDMs approximately $N$-representable, the optimization procedure did not bring the fixed 2-RDMs necessarily closer to the exact solution in this case.

	Our method is most suited for fixing 2-RDMs which are close to being $N$-representable, \textit{i.e.} the underlying theory gives 2-RDMs which do not severely violate the $N$-representability conditions. By using the $N$-representability optimization procedure, a small correction is added to the initial 2-RDM such that the final 2-RDM is $N$-representable in the sense that it has the right symmetry, fulfills the trace conditions and the $P$-, $Q$- and $G$-conditions.
	
	\begin{acknowledgements}
	D.V.N., S.D.B. and K.G. acknowledge support from the Research Foundation Flanders (FWO Vlaanderen). C.L. and P.W.A. thank Sharcnet and Compute Canada for computational resources.	
	\end{acknowledgements}

	\appendix*
	\section{Proofs}
	Higham\cite{higham1988sdp} showed that for an arbitrary matrix $B_0$ there is a unique nearest, with respect to the Frobenius norm, symmetric positive semidefinite matrix $B$, which can be expressed in terms of the eigenvalues and eigenvectors of the symmetric part of $B_0$. This matrix $B$ was named the positive approximant of $B_0$. The symmetric part of $B_0$ is given by
	\begin{equation}
		B_0^{sym} = \frac{1}{2} (B_0 + B_0^T)
	\end{equation}
	Since a positive approximant of $B_0$ is also a positive approximant of its symmetric part $B_0^{sym}$, the problem reduces to finding a solution to 
	\begin{equation} \label{app:higham}
		\begin{aligned}
			&\underset{B}{\text{minimize}}
			& & \| B - B_0^{sym}\|^2 \\
			& \text{subject to} & & B = B^T\\
			&&& B \, \succeq \, 0
		\end{aligned}
	\end{equation}
	Given the spectral decomposition for $B_0^{sym}$
	\begin{equation}
		B_{0}^{sym} = \sum_i \lambda_iZ_i^{}Z_i^T
	\end{equation}
	we can express the target matrix $B$ in the same eigenbasis:
	\begin{equation} 
		\label{eigB}
		B = \sum_{ij} Y_{ij}Z_i^{}Z_j^T
	\end{equation}
	The Frobenius difference can be rewritten as
	\begin{eqnarray}
		\|B - B_0^{sym}\|^2	&=& \sum_{ij} (Y_{ij} - \delta_{ij}\lambda_i)^2 \nonumber\\
		&=& \sum_{i \neq j} Y_{ij}^2 + \sum_{i} (Y_{ii} - \lambda_i)^2
	\end{eqnarray}
	For any positive semidefinite matrix $Y$ this cost function is minimized by setting the off-diagonal matrix elements to zero. Hence, the minimizing $B$ must be diagonal in the eigenbasis of $B_0^{sym}$. If we set $y_i^2 = Y_{ii}$, then
	\begin{equation}
		\|B - B_0^{sym}\|^2 = \sum_{\lambda_i \geq 0} (y_i^2 - \lambda_i)^2 + \sum_{\lambda_i < 0}(y_i^2 - \lambda_i)^2
	\end{equation}
	The last term is minimized when $y_i = 0$ and the first term is minimized when $y_i^2 = \lambda_i$. This means that the solution to the problem in Eq. \eqref{app:higham} is simply the positive part of $B_0^{sym}$, \textit{i.e.}
	\begin{equation}
		B = \sum_i \theta(\lambda_i)\lambda_iZ_i^{}Z_i^T
	\end{equation}
	with $\theta(\lambda)$ the Heaviside step function.
	
\begin{figure}
		\centering
		\includegraphics[scale=0.5]{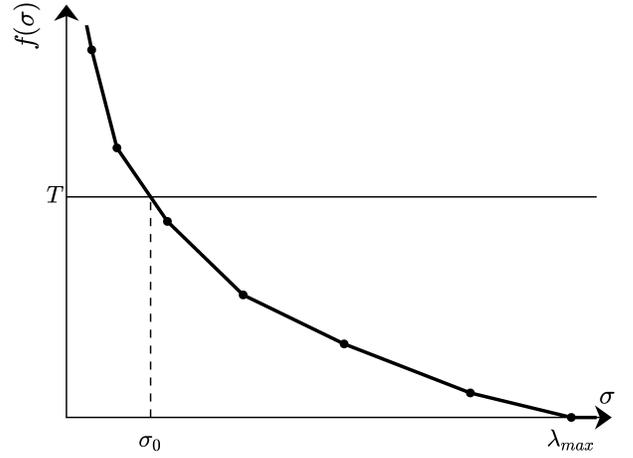}
		\caption{\label{fig:proof} $f(\sigma) = \sum_i \theta(\lambda_i - \sigma)(\lambda_i - \sigma)$.}
	\end{figure}

	Since the trace of a matrix is equal to the sum of its eigenvalues, it is easy to see that Higham's method does not preserve the normalization of the initial matrix $B_0$. Thus, we need to add another constraint in order to keep the trace fixed. Following the same reasoning as before and using the same definitions for $B_0^{sym}$ and $B$ we can again conclude that $Y$ should be a diagonal matrix. This is because the constraint on the trace $\sum_i Y_{ii} = T$ does not depend on the off-diagonal matrix elements. Setting $y_i^2 = Y_{ii}$ we can rewrite the problem as
	\begin{equation}
		\begin{aligned}
			&\underset{y_i}{\text{minimize}}
			& & \sum_i\left(\, y_i^2 - \lambda_i\right)^2 \\
			& \text{subject to} & & \sum_i y_i^2 = T
		\end{aligned}
	\end{equation}
	This is a constrained minimization problem which can be solved by the method of Lagrange multipliers. We introduce the Lagrangian with Lagrange multiplier $\mu$:
	\begin{equation}
		\mathcal{L}(\{y_j\}, \mu) = \sum_j (y_j^2 - \lambda_j)^2 - \mu \left(\sum_j y_j^2 - T\right)
	\end{equation}
	Taking the proper partial derivatives returns the following set of equations
	\begin{equation}
		\begin{cases}
			y_i\left[y_i^2 - \left(\lambda_i + \frac{\mu}{2}\right)\right] = 0 \\
			\sum_i y_i^2 = T
		\end{cases}
	\end{equation}
	This means either $y_i = 0$ or $y_i^2 = \lambda_i + \frac{\mu}{2}$, i.e. a constant shift $\sigma = -\frac{\mu}{2}$ from $\lambda_i$.
	Introducing the continuous piecewise linear function $f(\sigma) = \sum_i\theta(\lambda_i - \sigma) (\lambda_i - \sigma)$, it is clear that the roots of $f(\sigma) = T$ will yield the needed constant shift.
	It is easy to see that $f(\sigma)$ is monotonically decreasing in $\sigma$, reaching zero for $\sigma \geq \lambda_{max}$ and $f(\sigma) \rightarrow +\infty$ for $\sigma \rightarrow -\infty$. Hence, for every $T >0$ there will be a unique solution for the shift $\sigma_0$ (Fig. \ref{fig:proof}). The minimizing matrix $B$ is 
	\begin{equation}
		B = \sum_i\theta(\lambda_i - \sigma_0)(\lambda_i - \sigma_0)Z_iZ_i^T
	\end{equation}
	
	\bibliography{biblio}
\end{document}